# Calculations of Photonic Crystal Fibers by the Galerkin Method with Sine Functions without a Refractive Index Approximation

Elka Karakoleva, Blagovesta Zafirova[*] and Andrey Andreev

*Georgi Nadjakov Institute of Solid State Physics*
*Bulgarian Academy of Sciences,*
*72, Tsarigradsko Chaussee,Blvd., 1784 Sofia,*
*\*e-mail:* blzaf@issp.bas.bg

**Abstract.** Results from the calculation of the basic characteristics of the photonic crystal fiber with two rings of holes are presented by the approach which takes into account the exact distribution of the refractive index over the cross section of the photonic crystal fiber. Formulae are theoretically derived for the calculation of holes with arbitrary shapes by dividing the material within them into rotated at different angles rectangles.

.

## 1. INTRONUCTION

Photonic crystal fibers (*PCFs*) attract considerable attention since their waveguiding structure provides possibilities for creation of new or improved properties compared to that of the conventional optical fibers. They can be prepared in such a way that insures a single mode operation practically within the whole range of transparency of the material from which they are made (*endlessly single mode fibers*) at the same time with a very small change in the mode field diameter. The core diameter of the single mode fiber can be very high - 20 – 30 μm, which allows a transmission of high level optical power. Depending on the core diameter *PCF* can reveal very low or very high optical nonlinearity. Single mode fibers can be produced with unique dispersion properties – with an ultra flattened dispersion over very wide spectral range, a zero dispersion, realized at a chosen wavelength (even in the visible range), a zero dispersion in several chosen spectral ranges. The fibers can be easily prepared with very high birefringence – until one order higher than the birefringence, achieved by the standard anisotropic fibers by a suitable combination of the locations and the sizes of the holes [1-12].

The high refractive index contrast of the materials of the *PCF* requires full-vector methods [13] to model the fibers accurately. The most used numerical methods are a plane wave expansion method (*PWM*) [14-16], a localized function method (*LFM*) [17,18], a beam propagation method (*BPM*) [19,20], a finite-element method (*FEM*) [21,22], a finite-difference method in the time domain (*FDTD*) [23], a finite-difference method in the frequency domain (*FDFD*) [24-31], a source-model technique (*SMT*) [32,33,13], and a highly accurate semi-analytical multipole method (*MM*) [34,35]. A brief review of their merits and drawbacks is given in [24].

The approximation of the refractive index of the *PCF* is the major factor limiting the accuracy of the calculated data [13]. The description of the refractive index profile of the fine structure of interfaces between domains with high contrast in the refractive indices must be very precisely because the effective index of the mode critically depends on it. Another drawback [24] of the Galerkin method is the great number of double integrals which must be solved.

To overcome these drawbacks a development of the application of the Galerkin method for *PCF* is proposed. The main idea is presented in [36-38] together with results from a numerical calculation of the fiber with one ring of circular holes and a comparison of the values of the effective index of the fundamental mode received by it and another numerical methods. Here a numerical calculation is presented by the development for *PCF* with two rings of circular holes. More over, expressions are given for an analytical calculation of the double integrals of the elements of the matrices of the modes of the *PCF* with holes with square and rectangular shapes and *PCF* with holes with arbitrary shapes approximated by layers of parallel rectangles rotated at different angles.

The expressions are obtained by an approach which considerably reduced the number of the doubles integrals in the case of holes with symmetrical shapes. The main idea for the proposed approach is outlined in [36, 37]. To overcome the loss of accuracy and to reduce the calculation time, all integrals in the proposed development are analytically calculated in the case of holes with circular, square and rectangular shapes.

## 2. FORMULATION OF THE PROBLEM

A translationally invariant *PCF* consisting of $N_h$ holes located in an optical medium (a host medium) is considered. Monochromatic light with angular frequency $\omega$ and time dependence $\exp(i\omega t)$ propagates along *PCF* in the direction of the axis *z*. It is looked for transverse distributions of the electric and the magnetic fields of the mode with respect to a Cartesian coordinate system *xOy* with an origin at the lower left angle of a rectangular material domain with dimensions $L_x$ ($0 \leq x \leq L_x$) and $L_y$ ($0 \leq y \leq L_y$) comprising the cross-section of *PCF* with an arbitrary location of the holes. It is assumed that the electric and the magnetic fields are zero at the domain boundaries. With an appropriate choice of its dimensions this approximation is reasonable due to the abrupt drop of the fields of the guided modes outside the core.

The electric and the magnetic fields of the mode are solutions of the vector wave equations:

$$\nabla^2 \vec{E} + \nabla\left[\vec{E}\cdot\frac{\nabla n^2}{n^2}\right] + n^2 k^2 \vec{E} = 0 \quad (1) \qquad \nabla^2 \vec{H} + \left[\frac{\nabla n^2}{n^2}\times(\nabla\times\vec{H})\right] + n^2 k^2 \vec{H} = 0 \quad (2)$$

In the Cartesian coordinate system the vector equations are decomposed into x, y and z components, where the fact that $\partial n/\partial z = 0$ is used. The dependence of the components of the fields on the coordinate z is assumed to be $\exp(i\beta z)$, where $\beta$ is the longitudinal constant of propagation.

It is looked for solutions of the transverse components of the electric and magnetic fields in the form:

$$E_x(x,y) = \sum_{\mu=1}^{\infty}\sum_{\nu=1}^{\infty} A_{\mu\nu}^{E}\Phi_{\mu\nu}(x,y) \quad (3) \qquad E_y(x,y) = \sum_{\mu=1}^{\infty}\sum_{\nu=1}^{\infty} B_{\mu\nu}^{E}\Phi_{\mu\nu}(x,y) \quad (4)$$

$$H_y(x,y) = \sum_{\mu=1}^{\infty}\sum_{\nu=1}^{\infty} A_{\mu\nu}^{H}\Phi_{\mu\nu}(x,y) \quad (5) \qquad H_x(x,y) = \sum_{\mu=1}^{\infty}\sum_{\nu=1}^{\infty} B_{\mu\nu}^{H}\Phi_{\mu\nu}(x,y) \quad (6)$$

where
$$\Phi_{\mu\nu}(x,y) = \left[2/(L_x L_y)^{1/2}\right]\sin(\sigma_\mu x)\sin(\rho_\nu y) \quad (7)$$

is a complete orthonormal set of sine functions which are orthogonal over the finite rectangular domain:

$$\int_0^{L_x} dx \int_0^{L_y} dy\, \Phi_{\mu\nu}(x,y)\Phi_{\mu'\nu'}(x,y) = \delta_{\mu\mu'}\delta_{\nu\nu'}, \quad (8)$$

$\sigma_\mu = (\mu\pi/L_x);\ \rho_\nu = (\nu\pi/L_y);\ \mu,\nu$ are integers; $A_{\mu\nu}^{E}, B_{\mu\nu}^{E}, A_{\mu\nu}^{H}, B_{\mu\nu}^{H}$ are unknown coefficients in the expansions of the $E_x$, $E_y$, $H_y$ and $H_x$ respectively.

Using the Galerkin method, the two systems of two partial differential equations are converted into two systems each of $2m_x m_y$ coupled linear algebraic equations ($m_x$ and $m_y$ are the numbers of members in the truncated sums over $\mu$ and $\nu$) for the unknown coefficients $A_{\mu\nu}^{E}, B_{\mu\nu}^{E}, A_{\mu\nu}^{H}, B_{\mu\nu}^{H}$ [39]:

$$\sum_{\mu=1}^{m_x}\sum_{\nu=1}^{m_y}(M_{\mu'\nu',\mu\nu}^{E} A_{\mu\nu}^{E} + N_{\mu'\nu',\mu\nu}^{E} B_{\mu\nu}^{E}) = (\beta/k)^2 A_{\mu'\nu'}^{E} \quad (9)$$

$$\sum_{\mu=1}^{m_x}\sum_{\nu=1}^{m_y}(R_{\mu'\nu',\mu\nu}^{E} A_{\mu\nu}^{E} + S_{\mu'\nu',\mu\nu}^{E} B_{\mu\nu}^{E}) = (\beta/k)^2 B_{\mu'\nu'}^{E} \quad (10)$$

$\mu' = 1,2,...,m_x, \qquad \nu' = 1,2,...,m_y$

$$\sum_{\mu=1}^{m_x}\sum_{\nu=1}^{m_y}(M^H_{\mu'\nu',\mu\nu}A^H_{\mu\nu} + N^H_{\mu'\nu',\mu\nu}B^H_{\mu\nu}) = (\beta/k)^2 A^H_{\mu'\nu'} \qquad (11)$$

$$\sum_{\mu=1}^{m_x}\sum_{\nu=1}^{m_y}(R^H_{\mu'\nu',\mu\nu}A^H_{\mu\nu} + S^H_{\mu'\nu',\mu\nu}B^H_{\mu\nu}) = (\beta/k)^2 B^H_{\mu'\nu'} \qquad (12)$$

$$\mu' = 1, 2, ..., m_x, \qquad \nu' = 1, 2, ..., m_y$$

where:

$$M^E_{\mu'\nu',\mu\nu} = \frac{4}{S}\int_0^{L_x}dx\int_0^{L_y}dy\left[(n^2 - n^2_{\mu\nu})P_{ssss} + 2\frac{\sigma_{\mu'}}{k^2}\ln(n)(\sigma_\mu P_{ccss} - \sigma_{\mu'}P_{ssss})\right] \qquad (13)$$

$$N^E_{\mu'\nu',\mu\nu} = \frac{8}{S}\frac{\sigma_{\mu'}}{k^2}\int_0^{L_x}dx\int_0^{L_y}dy\,\ln(n)(\rho_\nu P_{sccs} + \rho_{\nu'}P_{scsc}) \qquad (14)$$

$$R^E_{\mu'\nu',\mu\nu} = \frac{8}{S}\frac{\rho_{\nu'}}{k^2}\int_0^{L_x}dx\int_0^{L_y}dy\,\ln(n)(\sigma_\mu P_{cssc} + \sigma_{\mu'}P_{scsc}) \qquad (15)$$

$$S^E_{\mu'\nu',\mu\nu} = \frac{4}{S}\int_0^{L_x}dx\int_0^{L_y}dy\left[(n^2 - n^2_{\mu\nu})P_{ssss} + 2\frac{\rho_{\nu'}}{k^2}\ln(n)(\rho_\nu P_{sscc} - \rho_{\nu'}P_{ssss})\right] \qquad (16)$$

$$M^H_{\mu'\nu',\mu\nu} = \frac{4}{S}\int_0^{L_x}dx\int_0^{L_y}dy\left[(n^2 - n^2_{\mu\nu})P_{ssss} + 2\frac{\sigma_\mu}{k^2}\ln(n)(\sigma_{\mu'}P_{ccss} - \sigma_\mu P_{ssss})\right] \qquad (17)$$

$$N^H_{\mu'\nu',\mu\nu} = -\frac{8}{S}\frac{\rho_\nu}{k^2}\int_0^{L_x}dx\int_0^{L_y}dy\,\ln(n)(\sigma_\mu P_{cscs} + \sigma_{\mu'}P_{sccs}) \qquad (18)$$

$$R^H_{\mu'\nu',\mu\nu} = -\frac{8}{S}\frac{\sigma_\mu}{k^2}\int_0^{L_x}dx\int_0^{L_y}dy\,\ln(n)(\rho_\nu P_{cscs} + \rho_{\nu'}P_{cssc}) \qquad (19)$$

$$S^H_{\mu'\nu',\mu\nu} = \frac{4}{S}\int_0^{L_x}dx\int_0^{L_y}dy\left[(n^2 - n^2_{\mu\nu})P_{ssss} + 2\frac{\rho_\nu}{k^2}\ln(n)(\rho_{\nu'}P_{sscc} - \rho_\nu P_{ssss})\right] \qquad (20)$$

$$n \equiv n(x,y) = \begin{cases} n_i = const, & x,y \in S_i \\ n_{host} = const, & x,y \notin S_i \end{cases}, \quad i = 1,2,...,N_h$$

$n_i$ is the constant refractive index of the ith hole with the surface $S_i$, $n_{host}$ is the constant refractive index of the host medium, $n^2_{\mu\nu} \equiv (\sigma_\mu^2 + \rho_\nu^2)/k^2$ is a dimensionless quantity,

$P_{ssss} \equiv P_{ssss}(x,y) = \sin(\sigma_\mu x)\sin(\sigma_{\mu'}x)\sin(\rho_\nu y)\sin(\rho_{\nu'}y)$,
$P_{ccss} \equiv P_{ccss}(x,y) = \cos(\sigma_\mu x)\cos(\sigma_{\mu'}x)\sin(\rho_\nu y)\sin(\rho_{\nu'}y)$.

The definitions of the remaining products are analogous.

Here the main idea of the development is briefly presented. Each of the integrals in (13-20) is a sum of double integrals over the host medium and over the holes in it. In (13-20) integrals over the holes surfaces are added and subtracted in which the refractive indices are replaced by the refractive index of the host medium in order not to integrate over the host medium:

$$\int_0^{L_x}\int_0^{L_y} f(x,y;n(x,y))dxdy = \int_0^{L_x}\int_0^{L_y} f(x,y;n_{host})dxdy + \sum_{i=1}^{N_h}\iint_{S_i}\left[f(x,y;n_i) - f(x,y;n_{host})\right]dxdy \quad (21)$$

e.g. the double integral over the domain with $N_h$ interfaces is replaced by a sum of a double integral over a homogeneous medium with a surface $S$ and a refractive index $n_{host}$ (where the orthogonality of the sine functions can be used) and $N_h$ homogeneous media with surfaces $S_i$, $i=1, 2,\ldots, N_h$ with changed refractive indices. That gives a possibility for the exact refractive index profile of the cross section of the *PCF* to be taken into account and for the double integrals to be analytically calculated in the case of circular, square and rectangular shapes of the holes.

Then the expressions (13-20) can be written as:

$$M^E_{\mu'\nu',\mu\nu} = \frac{4}{S}\sum_{i=1}^{N_h+1}\left[n_s^i I_{ssss}^i + 2\frac{\sigma_{\mu'}}{k^2}\ln(n_d^i)(\sigma_\mu I_{ccss}^i - \sigma_{\mu'} I_{ssss}^i)\right] \quad (22)$$

$$N^E_{\mu'\nu',\mu\nu} = \frac{8}{S}\sum_{i=1}^{N_h+1}\frac{\sigma_{\mu'}}{k^2}\ln(n_d^i)(\rho_\nu I_{sccs}^i + \rho_{\nu'} I_{scsc}^i) \quad (23)$$

$$R^E_{\mu'\nu',\mu\nu} = \frac{8}{S}\sum_{i=1}^{N_h+1}\frac{\rho_{\nu'}}{k^2}\ln(n_d^i)(\sigma_\mu I_{cssc}^i + \sigma_{\mu'} I_{scsc}^i) \quad (24)$$

$$S^E_{\mu'\nu',\mu\nu} = \frac{4}{S}\sum_{i=1}^{N_h+1}\left[n_s^i I_{ssss}^i + 2\frac{\rho_{\nu'}}{k^2}\ln(n_d^i)(\rho_\nu I_{sscc}^i - \rho_{\nu'} I_{ssss}^i)\right] \quad (25)$$

$$M^H_{\mu'\nu',\mu\nu} = \frac{4}{S}\sum_{i=1}^{N_h+1}\left[n_s^i I_{ssss}^i + 2\frac{\sigma_\mu}{k^2}\ln(n_d^i)(\sigma_{\mu'} I_{ccss}^i - \sigma_\mu I_{ssss}^i)\right] \quad (26)$$

$$N^H_{\mu'\nu',\mu\nu} = -\frac{8}{S}\sum_{i=1}^{N_h+1}\frac{\rho_\nu}{k^2}\ln(n_d^i)(\sigma_\mu I_{cscs}^i + \sigma_{\mu'} I_{sccs}^i) \quad (27)$$

$$R^H_{\mu'\nu',\mu\nu} = -\frac{8}{S}\sum_{i=1}^{N_h+1}\frac{\sigma_\mu}{k^2}\ln(n_d^i)(\rho_\nu I_{cscs}^i + \rho_{\nu'} I_{cssc}^i) \quad (28)$$

$$S^H_{\mu'\nu',\mu\nu} = \frac{4}{S}\sum_{i=1}^{N_h+1}\left[n_s^i I_{ssss}^i + 2\frac{\rho_\nu}{k^2}\ln(n_d^i)(\rho_{\nu'} I_{sscc}^i - \rho_\nu I_{ssss}^i)\right] \quad (29)$$

where:

$$n_s^i = \begin{cases} n_i^2 - n_{host}^2, & i=1,2,\ldots,N_h \\ n_{host}^2 - n_{\mu\nu}^2, & i = N_h+1 \end{cases} ; \quad n_d^i = \begin{cases} n_i/n_{host}, & i=1,2,\ldots,N_h \\ n_{host}, & i = N_h+1 \end{cases} ;$$

$$I_{ssss}^i = \begin{cases} I_{ssss}^{S_i} = \iint_{S_i} P_{ssss}(x,y)dxdy & i=1,2,\ldots,N_h \\ I_{ssss}^S = \iint_S P_{ssss}(x,y)dxdy & i = N_h+1 \end{cases}$$

The number $N_h+1$ is referred to the material domain. The definitions of the remaining integrals are analogous.

Let us consider the ith hole. A local coordinate system is introduced with an origin at the centre of the ith hole and axes parallel to the axes of the global coordinate system *xOy* (Fig. 1) in the case of a circular hole and axes rotated at an angle $\theta_i$ with respect to the axe *x* of the global coordinate system (Fig. 2) in the case of a rectangular hole. It was shown in [36,37] that in the case of circular holes the double integrals in (22-29) can be reduced to four integrals $I_{mm}$, $I_{mp}$, $I_{pm}$ and $I_{pp}$:

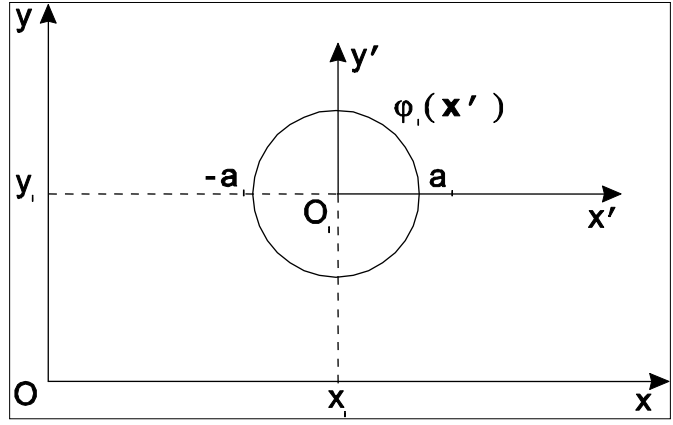

**FIGURE 1.** A Local Coordinate System $x'O_iy'$ with an Origin at the Centre of the ith Hole of the *PCF* and Axes Parallel to the Axes of the Global Coordinate System $xOy$.

$$I^{S_i}_{ssss} = \left( c_{mm}I_{mm} - c_{mp}I_{mp} - c_{pm}I_{pm} + c_{pp}I_{pp} \right)/4$$

$$I^{S_i}_{ccss} = \left( c_{mm}I_{mm} - c_{mp}I_{mp} + c_{pm}I_{pm} - c_{pp}I_{pp} \right)/4$$

$$I^{S_i}_{sscc} = \left( c_{mm}I_{mm} + c_{mp}I_{mp} - c_{pm}I_{pm} - c_{pp}I_{pp} \right)/4$$

$$I^{S_i}_{sccs} = \left( -s_{mm}I_{mm} + s_{mp}I_{mp} - s_{pm}I_{pm} + s_{pp}I_{pp} \right)/4$$

$$I^{S_i}_{scsc} = \left( s_{mm}I_{mm} + s_{mp}I_{mp} + s_{pm}I_{pm} + s_{pp}I_{pp} \right)/4$$

$$I^{S_i}_{cssc} = \left( -s_{mm}I_{mm} - s_{mp}I_{mp} + s_{pm}I_{pm} + s_{pp}I_{pp} \right)/4$$

$$I^{S_i}_{cscs} = \left( s_{mm}I_{mm} - s_{mp}I_{mp} - s_{pm}I_{pm} + s_{pp}I_{pp} \right)/4$$

where $c_{jk} = \cos(\sigma_j x_i)\cos(\rho_k y_i)$, $s_{jk} = \sin(\sigma_j x_i)\sin(\rho_k y_i)$ are "address" functions depending only on the location of the holes with respect to the global coordinate system and

$$I_{jk} = \int_{-a_i}^{a_i} \int_{-\varphi_i(x')}^{\varphi_i(x')} dx'dy' \cos(\sigma_j x')\cos(\rho_k y') \qquad j,k = m, p$$

are integrals which depend on the shape of the hole, but not on its location. Here $x_i$ and $y_i$ are the coordinates of the centre of the ith hole in the global coordinate system, $a_i$ is its radius, $\varphi_i(x') = \sqrt{a_i^2 - (x')^2}$ and $\sigma_m \equiv \sigma_\mu - \sigma_{\mu'}$; $\sigma_p \equiv \sigma_\mu + \sigma_{\mu'}$; $\rho_m \equiv \rho_\nu - \rho_{\nu'}$; $\rho_p \equiv \rho_\nu + \rho_{\nu'}$. For all holes with identical shapes the four integrals can be solved only once. For completeness the expressions are given for an analytical calculation of the four integrals for circular holes:

$$I_{jk} = \int_{-a_i}^{a_i} \int_{-\sqrt{a_i^2 - (x')^2}}^{\sqrt{a_i^2 - (x')^2}} dx'dy' \cos(\sigma_j x' + \rho_k y') = \frac{2\pi a_i}{\sqrt{\sigma_j^2 + \rho_k^2}} J_1\left(a_i\sqrt{\sigma_j^2 + \rho_k^2}\right) \quad j,k = m,p,$$

where $J_1$ is the Bessel function of order 1. In the case of *PCF* with circular holes all integrals in the matrices elements are analytically solved.

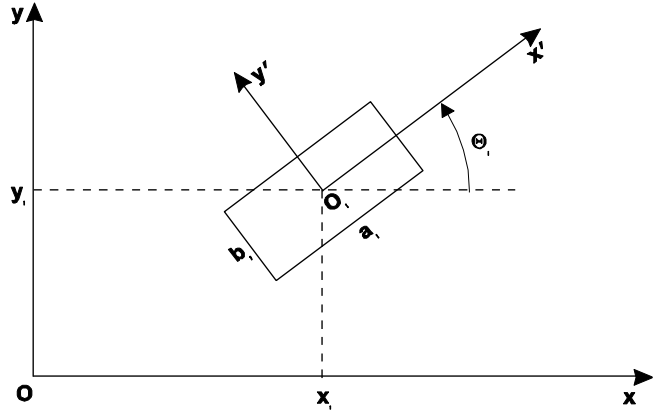

**FIGURE 2.** A Local Coordinate System $x'O_i y'$ with an Origin at the Centre of the ith Hole with a Rectangular Shape and Axes Rotated at an Angle $\theta_i$ with Respect to the Axe $x$ of the Global Coordinate System.

When the hole is with square or rectangular shape or the arbitrary shape of the hole is approximated by a layer of parallel rectangles, rotated at an angle $\theta_i$, then the double integrals in (22-29) are calculated by the following formulae:

$$I^{S_i}_{ssss} = \left( u_{mm} - u_{mp} - u_{pm} + u_{pp} \right)/2$$

$$I^{S_i}_{ccss} = \left( u_{mm} - u_{mp} + u_{pm} - u_{pp} \right)/2$$

$$I^{S_i}_{sscc} = \left( u_{mm} + u_{mp} - u_{pm} - u_{pp} \right)/2$$

$$I^{S_i}_{sccs} = \left( -v_{mm} + v_{mp} - v_{pm} + v_{pp} \right)/2$$

$$I^{S_i}_{scsc} = \left( v_{mm} + v_{mp} + v_{pm} + v_{pp} \right)/2$$

$$I^{S_i}_{cssc} = \left( -v_{mm} - v_{mp} + v_{pm} + v_{pp} \right)/2$$

$$I^{S_i}_{cscs} = \left( v_{mm} - v_{mp} - v_{pm} + v_{pp} \right)/2$$

$$u_{jk} = c^-_{jk} A^-_{jk} + c^+_{jk} A^+_{jk}$$

$$v_{jk} = c^-_{jk} A^-_{jk} - c^+_{jk} A^+_{jk}$$

$$c^{\pm}_{jk} = \cos(\sigma_j x_i \pm \rho_k y_i)$$

$$A^-_{jk} = \left[\sin(p^-_{jk} a_i/2)/p^-_{jk}\right]\left[\sin(q^+_{jk} b_i/2)/q^+_{jk}\right]$$

$$A^+_{jk} = \left[\sin(p^+_{jk} a_i/2)/p^+_{jk}\right]\left[\sin(q^-_{jk} b_i/2)/q^-_{jk}\right]$$

$$p^{\pm}_{jk} = \sigma_j \cos\theta_i \pm \rho_k \sin\theta_i$$

$$q^{\pm}_{jk} = \sigma_j \sin\theta_i \pm \rho_k \cos\theta_i$$

$$j,k = m,p.$$

Here $i$ is the successive number of the rectangle with dimensions $a_i$ and $b_i$, rotated at an angle $\theta_i$ with respect to the axe $x$ of the global coordinate system. When the holes with arbitrary shapes are approximated by layers of parallel rectangles all integrals are analytically solved with except of those referring to the rectangles comprising the irregular ends of the holes. There an averaging can be made over the parts of faces occupied by materials with different refractive indices before beginning of the calculations and the integrals can be analytically calculated for them. The systems of algebraic equations (9-12) are written in the form of matrix eigenvalue equations: $\hat{C}^E \vec{X}^E = (\beta/k)^2 \vec{X}^E$; $\hat{C}^H \vec{X}^H = (\beta/k)^2 \vec{X}^H$, where

$$\hat{C}^E \equiv \begin{pmatrix} \hat{M}^E & \hat{N}^E \\ \hat{R}^E & \hat{S}^E \end{pmatrix}; \quad \hat{C}^H \equiv \begin{pmatrix} \hat{M}^H & \hat{N}^H \\ \hat{R}^H & \hat{S}^H \end{pmatrix}; \quad \hat{M}^E, \hat{N}^E, \hat{R}^E, \hat{S}^E; \hat{M}^H, \hat{N}^H, \hat{R}^H, \hat{S}^H$$

are matrices consisting of the coefficients $M^E_{\mu'\nu',\mu\nu}; N^E_{\mu'\nu',\mu\nu}; R^E_{\mu'\nu',\mu\nu}; S^E_{\mu'\nu',\mu\nu}; M^H_{\mu'\nu',\mu\nu}; N^H_{\mu'\nu',\mu\nu}; R^H_{\mu'\nu',\mu\nu}; S^H_{\mu'\nu',\mu\nu}; \vec{X}^E = (\vec{A}^E, \vec{B}^E)^T$ and $\vec{X}^H = (\vec{A}^H, \vec{B}^H)^T$ are eigenvectors, consisting of the unknown coefficients $A^E_{\mu\nu}; B^E_{\mu\nu}; A^H_{\mu\nu}; B^H_{\mu\nu}$ and $(\beta/k)^2$ are unknown eigenvalues. The method is reduced to a suitability for coding and is incorporated into a created single *Visual FORTRAN 6.5* code. It calculates matrices elements, modal effective indices and transverse components of both the electric and magnetic fields propagating along the *PCF*. *EISPACK* [40] is incorporated into the code and is used to solve the eigenvalue equations.

## 3. NUMERICAL RESULTS

The *PCF* under consideration consists of two rings of cylindrical air holes each with a diameter $d = 5.0$ μm and a refractive index $n_i = 1.0$ ($i = 1,2,…,18$), arranged in a hexagonal lattice with a constant (a pitch) $\Lambda = 6.75$ μm within a host medium with a refractive index $n_{host} = 1.45$. The wavelength is $\lambda = 1.45$ μm. The vector distribution of the transverse magnetic field $\vec{H}_t(x, y) = \vec{H}_x(x, y) + \vec{H}_y(x, y)$ and the contour map of the magnetic field of the fundamental mode of the *PCF* with two rings of circular holes are shown on Figure 3 and Figure 4:

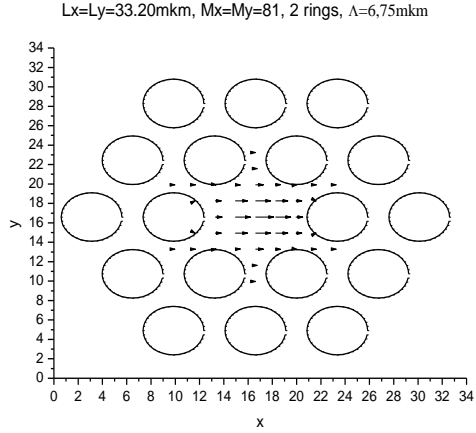

**FIGURE 3.** The Vector Distribution of the Transverse Magnetic Field $\vec{H}_t(x, y)$ (Linearly Polarized Along the Axe *x*) of the Fundamental Mode over the Cross Section of the *PCF* with Two Rings of Holes.

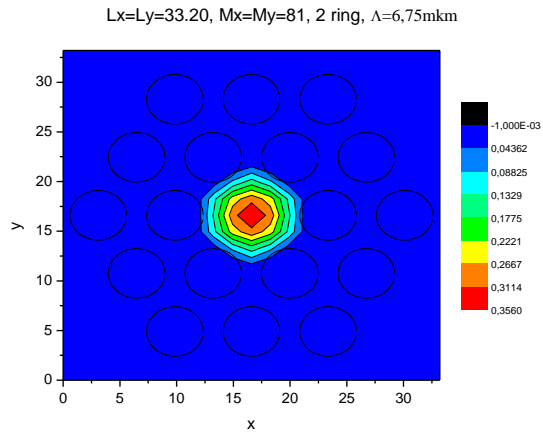

**FIGURE 4.** A Contour Map of the Magnetic Field of the Fundamental Mode of the *PCF* with Two Rings of Holes.

The essential part of the results is shown on *Table 1*. It can be seen that the value of $n_{eff}$ with smallest relative errors $\Delta_m = 3.1459 \times 10^{-8}$ and $\Delta_L = -6.0466 \times 10^{-9}$ is a solution of the problem, $n_{eff} = 1.44539622$. Here $\Delta_m$ is the smallest relative error between two successive solutions with terms in their expansions which differs by 1 and $\Delta_L$ is the smallest relative error for two solutions at two successive values of the dimensions of the material domain which differs by 0.2 μm.

**TABLE 1.** The Convergence of the Solution for $n_{eff}$ of the Fundamental Mode of the *PCF* with Two Rings of Holes.

| $L_x=L_y$ [μm] | $m_x=m_y$ | $n_{eff}$ | $\Delta_m$ | $\Delta_L$ |
|---|---|---|---|---|
| 32.20 | 79 | 1.445396232634340 | 1.1757E-08 | - |
| 32.40 | 81 | 1.445396308797130 | 2.1309E-08 | 5.2693E-08 |
| 32.60 | 80 | 1.445396265103050 | 3.5066E-08 | -3.0230E-08 |
| 32.80 | 81 | 1.445396247724620 | 7.7172E-10 | -1.2023E-08 |
| 33.00 | 81 | 1.445396233574110 | 1.0139E-08 | -9.7901E-09 |
| 33.20 | 81 | 1.445396224834410 | 3.1459E-08 | -6.0466E-09 |
| 33.40 | 82 | 1.445396265177670 | 3.4329E-08 | 2.7912E-08 |
| 33.60 | 82 | 1.445396247763510 | 3.2795E-08 | -1.2048E-08 |
| 33.80 | 82 | 1.445396221895270 | 3.2206E-08 | -1.7897E-08 |
| 34.00 | 80 | 1.445395906427260 | 4.0761E-08 | -2.1826E-07 |
| 34.20 | 80 | 1.445395820298730 | 4.1892E-08 | -5.9588E-08 |

The three dimensional graphics of the magnetic field of the fundamental mode of the *PCF* with two rings of holes is shown on Figure 5.

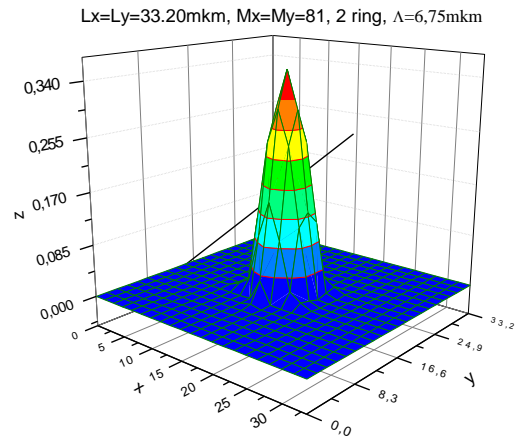

**FIGURE 5.** The Three Dimensional Graphics of the Magnetic Field of the Fundamental Mode of the *PCF* with Two Rings of Holes.

The values of the effective refractive index of the fundamental mode of the *PCF* with one and two rings of holes, calculated by the proposed development, the *PWM* and the adjustable boundary condition method (*ABC*) are given in the Table 2.

**TABLE 2.** Comparison of the Values of the Effective Index of the Fundamental Mode of the *PCF* under Consideration

| Method | Number of the Rings | Reference | $n_{eff}$ | Number of the Members in the Expansions | Relative Error in the Solution Convergence | Number of the Retained Digits |
|---|---|---|---|---|---|---|
| *PWM* | 1 | [16] | 1.4453952 | 131072 | $10^{-6}$-$10^{-7}$ | |
| *ABC* | 1 | [41], [42] | 1.445397228 | | | 7 |
| *Our Results* | 1 | [38] | 1.44539725467 | 68 | $4\times10^{-8}$/$8\times10^{-12}$ | 7/11 |
| | 2 | | 1.44539622 | 81 | $3\times10^{-8}$/$-6\times10^{-9}$ | 7/8 |

The effective index of the fundamental mode of the fiber with one ring of circular holes retains 7 digits when the number of the members in the expansions is changed by 1 and 11 digits when the dimensions of the material domain are changed in the process of the solution convergence. For the fiber with two rings the retained digits are 7 and 8 respectively. Taking into account the exact profile of the refractive index of the fiber and analytically calculating all integrals (i.e. the matrices in the eigenvalue problems) the proposed development increases the accuracy of the calculation of the effective refractive index with one to two orders and reduces the number of the members in the expansions of the field with three orders in comparison with PWM.

The theoretical derived expressions for the holes with square or rectangular shapes are incorporated in the created code. Numerical calculations by this code both for entire holes and for parts of the holes will be the subject of future study.

## 4. CONCLUSION

It is presented a numerical calculation with the development of the Galerkin method in its application for modeling of *PCF*. The effective index of the fundamental mode of the fiber with two rings of holes with circular shape is calculated with a high accuracy: 7 digits retain when the number of the members in the expansion is changed by 1 and 8 digits retain when the dimensions of the material domain are changed in the process of the solution convergence. The relative error of the effective index of the fundamental mode of a *PCF* with two rings of holes is with less value than that of the effective index of the fundamental mode of a *PCF* with one ring of holes calculated by *PWM* with one to two orders and the number of the members of the expansions of the magnetic field in a *PCF* with two rings of holes is with less value than that in a *PCF* with one ring of holes used by *PWM* with three orders. It is shown also the vector distribution of the transverse magnetic field, its contour map and three-dimensional plot. The proposed expressions for an analytical calculation of the double integrals of the elements of the matrices of the modes of the *PCF* with holes with square or rectangular shapes and of the *PCF* with holes with arbitrary shapes approximated by layers of parallel rectangles rotated at different angles give a possibility for an accurate and fast calculation of the *PCF*. The high accuracy of the received results and the small number of members in the field expansion show that the proposed development can be successfully used for modeling of the *PCF*.